\documentclass[default]{aastex701}
\usepackage{amsmath}
\usepackage{subcaption}

\begin{document}

\title{When Stabilizing Feedbacks Wreak Havoc in Habitable Planets: Chaos at the Freezing Point in a Four-Feedback Climate Model}

\author[0009-0007-4584-4417]{Chaucer Langbert}
\affiliation{Lunar and Planetary Laboratory, The University of Arizona, Tucson, AZ 85721, USA}
\email[hide]{chaucer@arizona.edu}  

\author[0000-0003-3714-5855]{D\'aniel Apai}
\affiliation{Steward Observatory, The University of Arizona, Tucson, AZ 85721, USA}
\affiliation{Lunar and Planetary Laboratory, The University of Arizona, Tucson, AZ 85721, USA}
\affiliation{James C. Wyant College of Optical Sciences, The University of Arizona, AZ 85721, USA}
\email[hide]{apai@arizona.edu}  

\author[0000-0002-1226-3305]{Renu Malhotra}
\affiliation{Lunar and Planetary Laboratory, The University of Arizona, Tucson, AZ 85721, USA}
\email[hide]{malhotra@arizona.edu}

\begin{abstract}

The long-term habitability of Earth-like planets is governed by the balance of positive and negative climate feedbacks that regulate surface temperature and atmospheric composition. While some feedbacks, such as ice-albedo and silicate weathering, are thought to operate on many terrestrial planets, the number and strength of additional climate feedbacks may vary substantially from world to world. An important open question is whether and how the introduction of an additional climate feedback influences long-term climate evolution and complexity. To investigate this question, we extended a low-order climate model including outgoing longwave radiation, ice--albedo, and carbonate--silicate weathering, with an additional generalized feedback. Across 35,864 simulations, we characterized dynamical behavior using the largest Lyapunov exponent (LLE). While 10.4\% of explored parameter combinations exhibited positive LLEs, indicating chaos, the fraction rose to 25.6\% within the stabilizing, near-freezing regime. This suggests that chaotic behavior is preferentially concentrated where a strong stabilizing feedback operates in the same temperature regime as a strong destabilizing feedback, contrary to the expectation that additional negative feedbacks should increase climate stability. Increasing normalized volcanic outgassing shifts the chaotic regime toward lower stellar flux and partially suppresses chaos at high instellation.

\end{abstract}


\section{Introduction}

Climate feedbacks regulate the long-term evolution of terrestrial planetary climates and play a central role in determining planetary habitability. Stabilizing feedbacks tend to damp perturbations and promote equilibrium climate states, while destabilizing feedbacks amplify departures from equilibrium. On Earth, the combined action of outgoing longwave radiation (OLR), ice--albedo interactions, and carbonate--silicate weathering is thought to maintain long-term climate stability despite substantial changes in solar luminosity and internal forcing over geologic time \citep[e.g.,][]{Menou2015,Haqq-Misra2016,Arnscheidt2020}.

The interaction of these feedbacks can nevertheless produce a rich range of dynamical behaviors. Previous studies have identified stable warm and globally glaciated states, bistability, and self-sustained climate oscillations arising from coupled ice--albedo and carbon-cycle feedbacks \citep[e.g.,][]{Menou2015,Haqq-Misra2016,Arnscheidt2020}. More recently, \cite{Langbert2026} explored a generalized climate model that included a fourth feedback in addition to OLR, ice--albedo, and carbonate--silicate weathering, demonstrating that complex climate trajectories, including chaotic behavior, can emerge under some conditions, primarily with stabilizing fourth feedbacks acting near the freezing point of water and across a wide range of instellations. However, the regions of parameter space that favor chaos and the physical circumstances under which it arises remain poorly understood.

Several candidate planetary feedbacks are expected to become strongly nonlinear near thermodynamic phase transitions. The freezing point of water is especially important because sea-ice growth and retreat can alter planetary albedo and produce nonlinear threshold behavior \citep{Winton2008,Eisenman2009}. On Earth, ocean carbonate chemistry regulates the partitioning of carbon between the atmosphere and ocean, while continental and seafloor weathering provide long-term sinks for atmospheric CO$_2$ \citep{Walker1981,Zeebe2012,KrissansenTotton2018}. Such threshold behavior is common in planetary climate systems and may produce interactions between otherwise stabilizing feedbacks that are difficult to anticipate from equilibrium analyses alone. Throughout this work, we use the term \emph{fourth feedback} to denote a generalized climate mechanism that contributes an additional temperature-dependent forcing. The feedback is not intended to represent a single identified process, but rather serves as a proxy for potentially diverse physical, chemical, or biological mechanisms that may operate on terrestrial planets. In particular, processes associated with the liquid--ice transition of water are expected to exhibit strongly nonlinear behavior because the availability, transport, and storage of liquid water can change rapidly across the freezing threshold. Such transitions may naturally generate feedbacks with characteristic amplitudes and response timescales that differ substantially from those operating far from phase boundaries. 

A key consequence of strongly nonlinear feedback interactions is the possibility of deterministic chaos. Chaotic systems evolve according to deterministic equations but exhibit sensitive dependence on initial conditions, causing nearby trajectories to diverge exponentially and limiting long-term predictability \citep{Gleick1987}. Chaos has been identified in a variety of geophysical and astrophysical systems \citep[e.g.,][]{Lorenz1969,Ghil1987,Pierrehumbert2010}, yet its prevalence and origin in planetary climate models merits deeper study. In particular, it is unclear whether chaotic climate evolution is a generic consequence of the number of distinct feedbacks or whether it requires specific combinations of feedback strength, timescale, and activation temperature. Although general circulation models capture many climate processes absent from low-order models, they cannot practically simulate the geologic timescales associated with carbonate--silicate weathering. Long-term climate evolution is therefore typically studied using simplified dynamical models or hybrid approaches that couple GCM results to reduced-order models. The behaviors identified here can therefore serve as targets for future investigation with more comprehensive climate models.

Here we build upon the generalized four-feedback climate model introduced by \cite{Langbert2026}. The model extends the canonical framework of OLR, ice--albedo, and carbonate--silicate weathering feedbacks by including a fourth temperature-dependent feedback represented by an additional dynamical variable. The system is governed by three coupled ordinary differential equations describing the evolution of surface temperature, atmospheric CO$_2$ partial pressure, and the state of the fourth feedback. The principal control parameters are stellar flux ($S$), volcanic outgassing rate normalized by the ingassing due to silicate weathering ($V/W_0$), fourth-feedback strength ($c$), and fourth-feedback activation temperature ($T_f$), which together determine the relative importance and operating regime of the four climate feedbacks. 

While \citet{Langbert2026} demonstrated that chaotic climate evolution can arise within this framework, the circumstances under which chaos occurs remained unclear. Although chaotic solutions comprised only a minority of the explored parameter space, they appeared to cluster within a well-defined region associated with strong stabilizing feedbacks acting near the temperature range over which the ice--albedo feedback is most sensitive. This raises a fundamental question: what physical conditions favor the emergence of chaos, and why does chaotic behavior occur preferentially in this region of parameter space?

In the present work, we systematically explore the parameter space to identify the regions that produce chaotic behavior and to determine how the occurrence of chaos depends on planetary forcing and feedback properties. We characterize climate evolution using the largest Lyapunov exponent (LLE) \citep{Eckmann1985,Wolf1985} and complementary bifurcation analyses, which allow transitions between stable equilibria, periodic oscillations, and chaotic attractors to be identified. Counterintuitively, we find that chaos is strongly concentrated when a stabilizing fourth feedback activates near the temperature range over which the ice--albedo feedback acts most strongly. Increasing normalized volcanic outgassing systematically shifts this chaotic regime toward lower stellar flux and partially suppresses chaos at high instellation. These results suggest that intrinsically unpredictable climate evolution may arise preferentially from the interaction of multiple nonlinear feedbacks operating over the same temperature interval rather than being an inherent property of terrestrial climate systems.

\section{Methods} \label{sec:methods}

\subsection{Dynamical System}

We adopt a zero-dimensional climate framework derived from earlier studies of long-term planetary climate regulation \citep{Walker1981,Menou2015,Arnscheidt2020}. The baseline framework includes outgoing longwave radiation (OLR), ice-albedo \citep{Budyko1969,Sellers1969}, and carbonate-silicate weathering feedbacks \citep{Walker1981}. We retain the \citet{Walker1981} weathering parameterization to maintain consistency with previous three-feedback climate models and to isolate the dynamical consequences of introducing an additional climate feedback. While more physically detailed weathering formulations have since been developed \citep[e.g.,][]{Maher2014,Graham2020}, these parameterizations are more complex, whereas the modeling approach taken here is to develop the most idealized possible framework to study chaos due to an unknown fourth feedback. Thus, evaluating the sensitivity of chaotic climate behavior to alternative weathering parameterizations is beyond the scope of this initial study. \cite{Langbert2026} generalized the \citet{Walker1981} framework by introducing a fourth climate feedback, allowing the effects of additional stabilizing or destabilizing processes to be explored. The goal of a more generalized fourth feedback is to remain agnostic to the potential diversity of chemical and physical processes potentially present on Earth-like planets. Rather than representing a single identified physical process, the fourth feedback is intended as a generalized proxy for climate mechanisms that may become strongly nonlinear near specific thermodynamic thresholds. 

Adopting the four-feedback formulation, we systematically vary the fourth feedback's strength and equilibrium temperature to characterize its influence on long-term climate behavior. In this work, we focus particularly on candidate feedbacks that are the consequence of the presence of liquid water, sea ice, or ocean circulation near the freezing point of water. The system of coupled ordinary differential equations is described by Equations~\ref{model1}-\ref{model3} \citep{Langbert2026}.
\begin{figure*}[htb!]
\begin{align}
    C \dot{T} &= \frac{S}{4}(1 - \alpha(T)) - \frac{S_0}{4}(1 - \alpha_0) - a(T - T_0) + b \ln\left(\frac{P}{P_0}\right) + c f \label{model1} \\
    \dot{P} &= V - W(T) e^{k(T - T_0)} \left(\frac{P}{P_0}\right)^{\beta} \label{model2} \\
    \dot{f} &= -\gamma_f \left[ f - \kappa \tanh(\delta_f(T - T_f)) \right] \label{model3}
\end{align}
\end{figure*}

The climate state is described by the phase-space variables surface temperature ($T$), atmospheric CO$_2$ partial pressure ($P$), and the state of the generalized fourth feedback ($f$). The stellar flux ($S$), normalized volcanic outgassing rate ($V/W_0$), fourth feedback strength ($c$), and fourth-feedback equilibrium temperature ($T_f$) are treated as control parameters. Relative to previous three-feedback models that include OLR, ice--albedo, and carbonate--silicate weathering \citep[e.g.,][]{Menou2015,Haqq-Misra2016,Arnscheidt2020}, the present formulation introduces a third dynamical variable and an addition temperature-dependent feedback with independently adjustable strength and activation temperature. Following \citet{Langbert2026}, we use the hyperbolic tangent in Equation~\ref{model3} as a simple bounded representation of a threshold-like feedback that smoothly transitions between two limiting states; other nonlinear activation functions could shift the location or extent of chaotic regimes, but the key ingredient explored here is the overlap between a rapidly varying fourth feedback and the ice--albedo transition.

This extension permits interactions between multiple stabilizing and destabilizing feedbacks that are not captured in the three-feedback framework. Individual simulations therefore evolve through the three-dimensional phase space $(T,P,f)$, whereas the grid search explores the four-dimensional parameter space $(S,V/W_0,c,T_f)$. Unless otherwise stated, all integrations use the fiducial parameters in Table~\ref{tab:model_params}, while the grid search varies $S$, $V/W_0$, $c$, and $T_f$ over the listed ranges.

\begin{table}[htb!]
\centering
\small
\begin{tabular}{l c c l}
\hline
\textbf{Parameter} & \textbf{Value/range} & \textbf{Units} & \textbf{Description} \\
\hline
$S_0$ & 1365 & W m$^{-2}$ & Present-day solar constant \\
$T_0$ & 288 & K & Reference surface temperature \\
$P_0$ & $3\times10^{-4}$ & bar & Reference CO$_2$ partial pressure \\
$T_i$ & 273 & K & Ice--albedo transition temperature \\
$T_{\rm scale}$ & 5 & K & Width of ice--albedo transition \\
$\alpha_c,\alpha_w$ & 0.6, 0.2 & -- & Cold and warm-state albedo \\
$\alpha_0$ & 0.241 & -- & Reference planetary albedo \\
$a$ & 2.2 & W m$^{-2}$ K$^{-1}$ & OLR climate restoring coefficient \\
$b$ & 8.0 & W m$^{-2}$ & CO$_2$ radiative forcing coefficient \\
$C$ & $2\times10^8$ & J m$^{-2}$ K$^{-1}$ & Effective heat capacity \\
$W_w$ & 70 & bar Gyr$^{-1}$ & Maximum weathering rate \\
$W_0$ & 62.831 & bar Gyr$^{-1}$ & Reference weathering rate at ($T_0$, $P_0$) \\
$k$ & 0.1 & K$^{-1}$ & Weathering temperature sensitivity \\
$\beta$ & 0.5 & -- & Weathering CO$_2$ sensitivity exponent \\
$\kappa$ & 1.0 & -- & Maximum fourth-feedback amplitude \\
$\delta_f$ & 0.1 & K$^{-1}$ & Fourth-feedback transition sharpness \\
$\gamma_f$ & $10^{-3}$ & yr$^{-1}$ & Fourth-feedback relaxation rate \\
\hline
$S$ & 900--1500 & W m$^{-2}$ & Stellar flux \\
$V/W_0$ & 0.5, 1, 2, 5, 10 & -- & Normalized volcanic outgassing rate \\
$c$ & $-50$ to $+50$ & W m$^{-2}$ & Fourth-feedback strength \\
$T_f$ & 250--300 & K & Fourth-feedback equilibrium temperature \\
\hline
$d_0$ & $10^{-8}$ & -- & Initial trajectory separation for LLE calculation \\
$t_{\rm trans}$ & $5\times10^7$ & yr & Transient integration time before LLE calculation \\
$\Delta t$ & $5\times10^5$ & yr & Trajectory renormalization interval \\
$N$ & 25 & -- & Maximum number of renormalization steps \\
\hline
\end{tabular}
\caption{
Model parameters, explored grid ranges, and numerical parameters used for Lyapunov exponent calculations. Here $S_0$, $T_0$, and $P_0$ define the Earth-normalized reference climate; $T_i$ and $T_{\rm scale}$ control the ice--albedo transition; $W_w$ and $W_0$ set the weathering and volcanic outgassing normalization; $(c,T_f,\gamma_f)$ describe the strength, transition temperature, and relaxation rate of the generalized fourth feedback; and the final rows summarize the numerical parameters used to compute the largest Lyapunov exponent.
}
\label{tab:model_params}
\end{table}

\begin{figure*}[htb!]
    \centering
    \includegraphics[width=\linewidth]{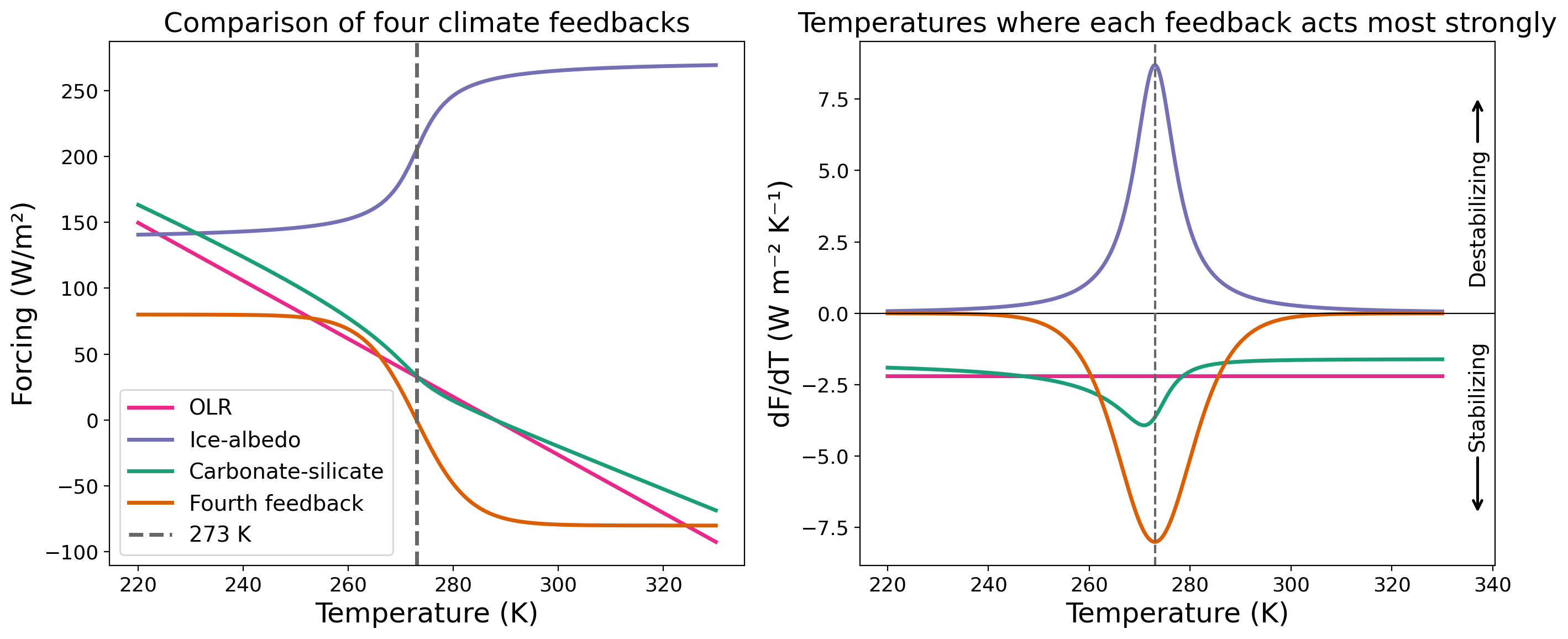}
    \caption{Comparison of the four climate feedbacks and their temperature dependence. Left: Forcing contributions versus temperature for OLR, ice--albedo, carbonate--silicate, and a representative fourth feedback. While the ice--albedo and fourth feedback both exhibit threshold-like behavior near the freezing point of water, they have the opposite dynamical effect: ice--albedo is destabilizing, whereas the fourth feedback acts to restore the climate toward equilibrium. Right: Temperature derivatives highlight where each feedback is strongest. The overlap between the destabilizing ice--albedo and a stabilizing fourth feedback near freezing illustrates the competing influences that characterize the chaotic regime identified in this work.}
    \label{fig:feedback_comparison}
\end{figure*}

In this work, we distinguish between feedbacks that are closely tied to the liquid--ice transition of water and feedbacks that operate more broadly across temperature space. Water-linked feedbacks are expected to exhibit strong nonlinear behavior near the freezing point because small temperature changes can produce large changes in ice cover, liquid-water availability, ocean circulation, and associated geochemical cycling. The generalized fourth feedback is intended to represent such threshold-sensitive processes without specifying a particular physical mechanism. Unlike the ice--albedo feedback, which is destabilizing and amplifies temperature perturbations, the fourth feedback considered here is typically stabilizing and acts to restore the climate toward its preferred equilibrium temperature. When the two feedbacks operate over similar temperature ranges, they compete: the ice--albedo feedback amplifies departures from equilibrium, while the fourth feedback opposes them. A comparison of the temperature dependence of the four climate feedbacks is shown in Figure~\ref{fig:feedback_comparison}.

The bump in the carbonate--silicate cycle curve near the freezing point of water, shown in the right panel of Figure~\ref{fig:feedback_comparison}, occurs because the weathering function turns on rapidly there, causing equilibrium atmospheric CO$_2$ and its associated greenhouse forcing to decrease sharply with temperature.

\subsection{Nullcline Analysis and Chaos Calculations}

To understand the equilibrium structure of the 3D climate system, we examined the nullclines of the coupled ODEs for temperature $T$, atmospheric CO$_2$ partial pressure $P$, and the fourth-feedback state variable $f$. A nullcline is the set of points in phase space where a single time derivative vanishes, irrespective of the values of the other derivatives. Thus, points on a given nullcline are not necessarily equilibrium solutions; rather, they identify locations where one state variable is momentarily stationary. Equilibrium points occur only at intersections of all three nullclines, where $\dot T = \dot P = \dot f = 0$.

The nullclines are obtained by setting each time derivative equal to zero. For the feedback variable, the condition $\dot f=0$ gives an explicit relation between $f$ and $T$,
\begin{equation}
f(T) = \kappa \tanh\!\left(\delta_f (T-T_f)\right).
\end{equation}
This defines the $f$-nullcline.

For the carbon cycle equation, setting $\dot P=0$ yields
\begin{equation}
V = W(T)\exp\!\left[k(T-T_0)\right]\left(\frac{P}{P_0}\right)^{\beta},
\end{equation}
which can be rearranged to give the $P$-nullcline as
\begin{equation}
P(T) = P_0 \left[\frac{V}{W(T)\exp\!\left(k(T-T_0)\right)}\right]^{1/\beta}.
\end{equation}

Finally, fixed points of the full system must also satisfy $\dot T=0$. Rather than solving simultaneously for $(T,P,f)$ in three dimensions, we substituted the two explicit nullcline relations above into the temperature equation, defining a reduced one-dimensional equilibrium condition along the intersection of the $P$- and $f$-nullclines:
\begin{equation}
    \left.\frac{dT}{dt}\right|_{P=P(T),\,f=f(T)} =
\frac{1}{C}\left[
\frac{S}{4}\left(1-\alpha(T)\right)
- \frac{S_0}{4}\left(1-\alpha_0\right)
- a(T-T_0)
+ b\ln\!\left(\frac{P(T)}{P_0}\right)
+ c\,f(T)
\right]. \label{Tnull} 
\end{equation}
Equilibria of the full 3D system are therefore given by the roots of
\begin{equation}
\left.\frac{dT}{dt}\right|_{P=P(T),\,f=f(T)} = 0.
\end{equation}
In practice, we evaluated this reduced function over a temperature grid and identified zero crossings to locate candidate equilibrium temperatures, then recovered the corresponding equilibrium values of $P$ and $f$ from $P(T)$ and $f(T)$.

\subsection{Largest Lyapunov Exponents (LLEs)}

In order to examine stability of the equilibria (that is, of the fixed points), we calculate the largest Lyapunov exponent (LLE). The LLE measures the exponential rate at which nearby trajectories in phase space diverge, providing a direct diagnostic of sensitive dependence on initial conditions. Positive values of the LLE indicate chaos, while negative values correspond to stable fixed points or limit cycles. For each simulation, we evolve the system forward in time with an implicit Backward Differentiation Formula (BDF) solver implemented in \texttt{scipy.integrate.solve\_ivp} \citep{Virtanen2020} until transients decay, ensuring that the trajectory samples the long-term attractor if any attractor exists.

The LLE is then estimated using a standard two-trajectory method. We initialize a perturbed copy of the reference trajectory with a small separation $d_0$ in a transformed state space defined by $(T, \log P, f)$, which regularizes the large dynamic range in pressure and places all variables on comparable scales. Both trajectories are integrated forward over successive time intervals, and after each interval the separation $d_1$ between them is measured. The perturbation vector is then renormalized to the original magnitude $d_0$ while preserving its direction, and the procedure is repeated. The LLE is computed as the time-averaged logarithmic growth rate,
\begin{equation}
    \lambda_{\max} = \frac{1}{N \, \Delta t} \sum_{i=1}^{N} \ln\left(\frac{d_1^{(i)}}{d_0}\right),
\end{equation}
where $N$ is the number of renormalization steps and $\Delta t$ is the integration interval. The parameter choices for subsequent LLE calculations are included in Table~\ref{tab:model_params}.

To ensure robust estimates, we require a minimum of five successful renormalization steps in the LLE calculation and exclude integrations that terminate prematurely due to excursions beyond the intended domain of validity of our simplified Earth-like climate parameterizations (temperatures outside 200–400 K or CO$_2$ partial pressures outside 10$^{-6}$–1000 bar), following \cite{Langbert2026}. Repeating the analysis with a narrower admissible temperature range (210–380 K) produced no qualitative changes in the prevalence or location of the chaotic regimes. 

We verified that the qualitative distribution of chaotic and non-chaotic solutions is robust to the numerical choices used in the LLE calculation. In particular, the sign and overall magnitude of the LLEs were insensitive to moderate variations in the initial perturbation amplitude $d_0$, renormalization interval $\Delta t$, integration duration, and random initial conditions. The transformed metric space $(T,\log P,f)$ was adopted to prevent the large dynamic range in CO$_2$ partial pressure from artificially dominating trajectory separations. Trajectories terminating at imposed physical bounds were excluded from the analysis.

\section{Results} \label{sec:results}

\subsection{Chaotic solutions in the parameter space}

We mapped the LLE across the four-dimensional parameter space $(S,V/W_0,c,T_f)$ to identify the conditions under which chaotic climate evolution occurs. The full grid results for all 35,864 simulations are shown in Figure~\ref{fig:grid search} in the Appendix.

Figure~\ref{fig:tf_trajectories} shows that, for an Earth-like climate configuration with present-day stellar flux ($S = S_\oplus$) and volcanic outgassing normalized to the present-day silicate weathering rate ($V/W_0 = 1$), the transition temperature $T_f$ of a stabilizing fourth feedback strongly influences the system's dynamical behavior. Here, $W_0$ is the model normalization corresponding to the present-day silicate weathering rate, which balances volcanic CO$_2$ outgassing under long-term carbon cycle equilibrium \citep{Walker1981,Berner2004}. As $T_f$ approaches the freezing point of water, the climate transitions from a stable equilibrium to a limit cycle and then to quasi-periodic or chaotic attractors (Figure~\ref{fig:tf_trajectories}). Small changes in activation temperature therefore lead to qualitatively different long-term climate evolution.

\begin{figure*}[htb!]
\centering
\includegraphics[width=\linewidth]{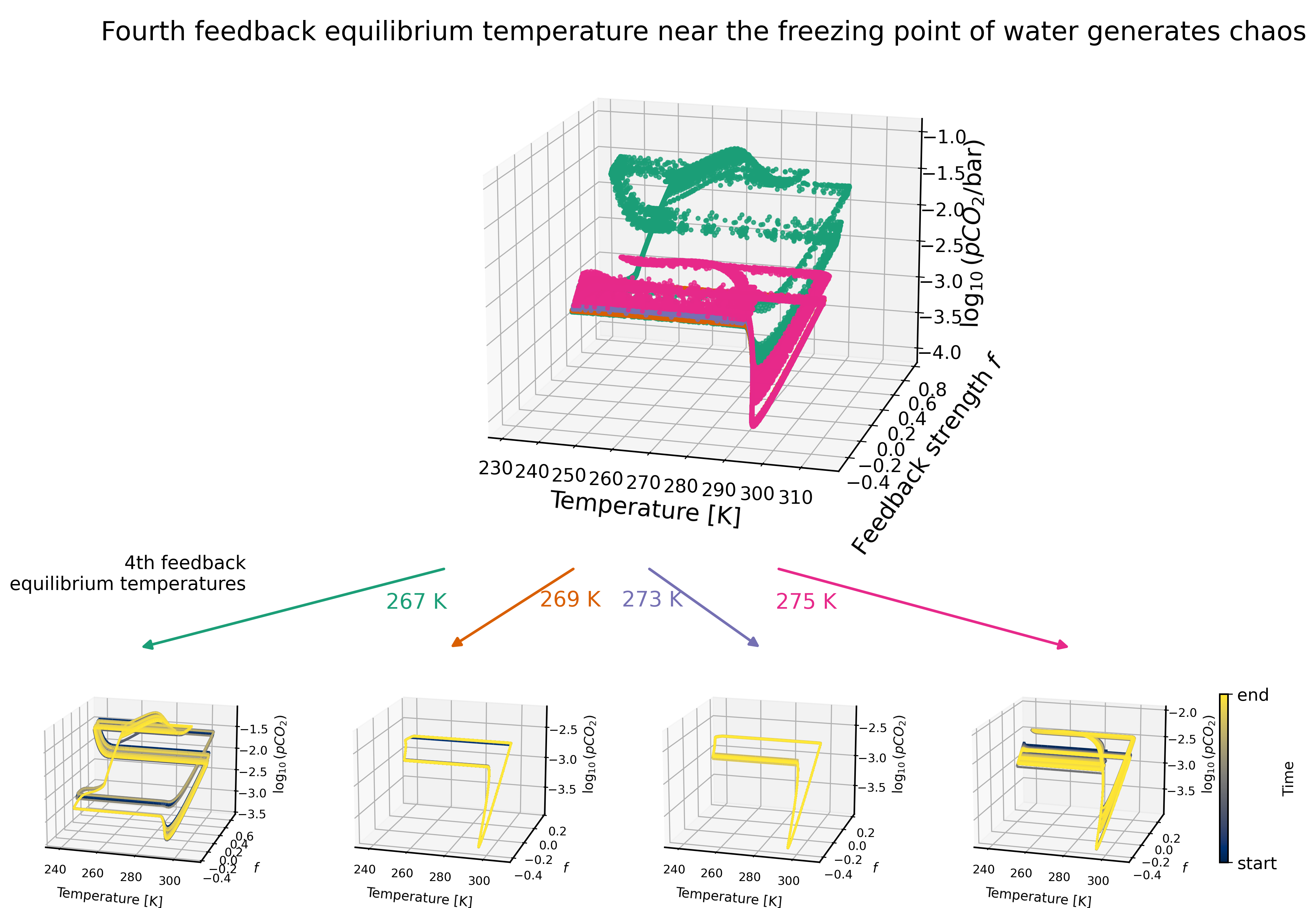}
\caption{Climate trajectories in $(T, f, \log_{10} p\mathrm{CO}_2)$ phase space for a fourth feedback with $c = -50$ W m$^{-2}$ and various transition temperatures $T_f$ (Earth-like $S$ and $V$). \textit{Top:} Overlaid trajectories for $T_f = 267$, 269, 273, and 275 K (100 Myr integrations after 400 Myr transients). \textit{Bottom:} Individual attractors showing chaos at 267 K and 275 K and limit cycles at 269--273 K. Small changes in the activation temperature near the freezing point, where the ice-albedo feedback acts most strongly, qualitatively alter the system dynamics, reflecting the strong sensitivity of the climate to feedback interactions in this regime.}
\label{fig:tf_trajectories}
\end{figure*}

Across the full parameter survey, chaotic solutions occupy a localized and contiguous region of parameter space. The strongest predictor of chaos is the combination of a strong stabilizing fourth feedback ($c \lesssim -30$ W m$^{-2}$) and an activation temperature near the ice--albedo transition ($T_f \approx 270$--282 K), where the two feedbacks operate over similar temperature ranges. Within this broad region, however, the occurrence of chaos also depends on stellar flux and normalized volcanic outgassing rate, as discussed below. Overall, 10.4\% of simulations are chaotic, but this fraction rises to 25.6\% within the near-freezing, strongly stabilizing regime and exceeds 75\% in some $(S, V/W_0)$ slices where $T_f \approx 270-282$ K and $c \lesssim -30$ W m$^{-2}$. Outside this region, solutions generally converge to stable fixed points or regular limit cycles.

We also examined fourth-feedback transition temperatures extending well above the freezing point but did not find comparably robust chaotic regimes associated with deglaciation. In our model, chaos remains localized near the freezing transition.

\subsection{Bifurcations and chaos at the freezing point}

\begin{figure*}[htb!]
\centering
\includegraphics[width=\linewidth,trim=1cm 18cm 1cm 18cm,clip]{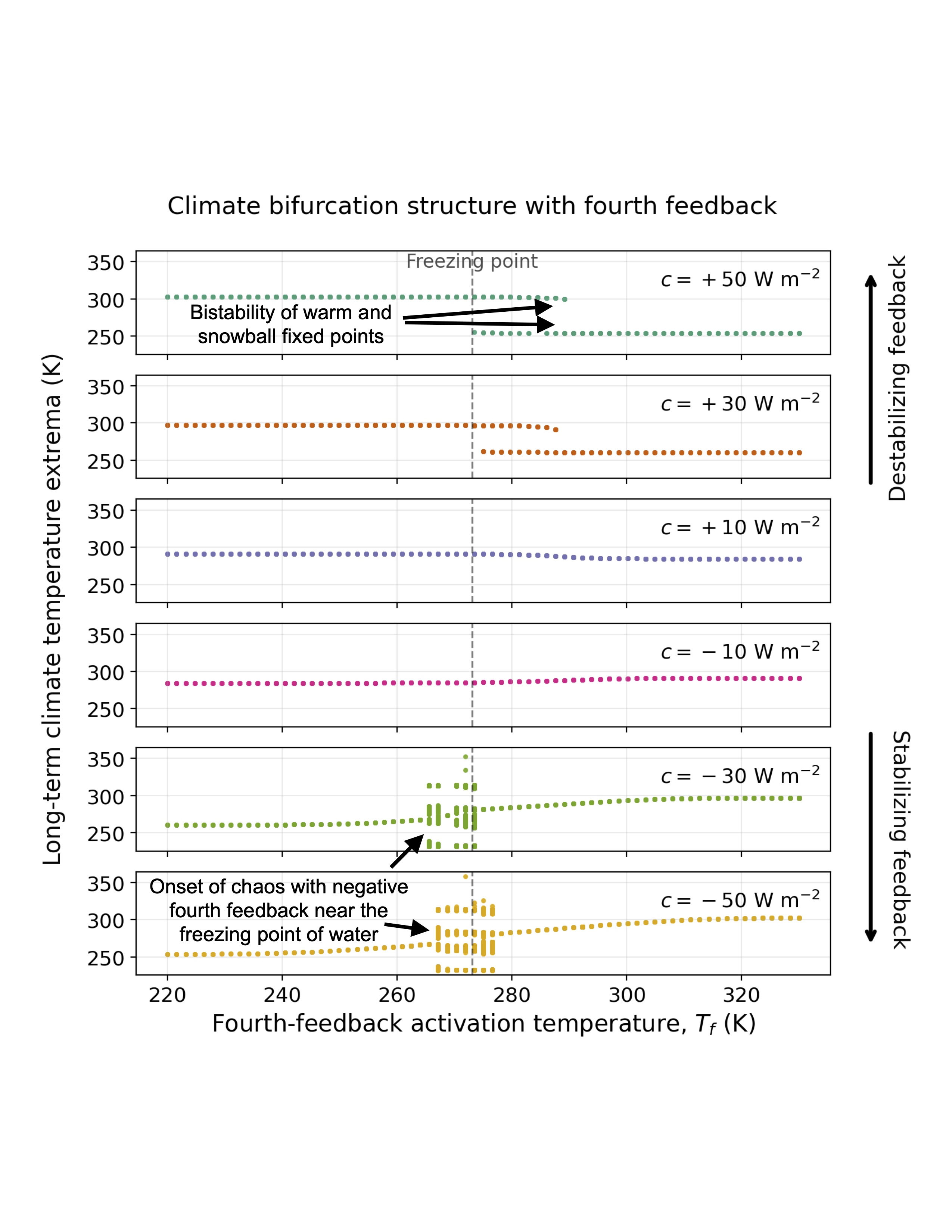}
\caption{
Bifurcation structure of the four-feedback climate model as a function of feedback transition temperature ($T_f$) and feedback strength ($c$). Each panel shows a sweep in $T_f$ for a fixed value of $c$. For each $(c,T_f)$ combination, the model was integrated from 10 random initial conditions generated independently for each parameter combination; the plotted points are the local temperature maxima and minima recorded after the transient evolution was removed. Single branches indicate convergence toward a stable equilibrium, whereas multiple extrema at the same $T_f$ indicate periodic or chaotic variability. The dashed vertical line marks the freezing point of water ($273$ K). Strong negative feedbacks ($c\lesssim -30$ W m$^{-2}$) generate complex dynamics when the feedback activates near the ice--albedo transition, while weaker feedbacks generally converge to stable climates. Strong positive feedbacks ($c=+30$ to $+50$ W m$^{-2}$) produce bistability between warm and snowball-like states.
}
\label{fig:bifn}
\end{figure*}

Figure~\ref{fig:bifn} shows the bifurcation structure of the four-feedback model as a function of feedback strength $c$ and transition temperature $T_f$. For each parameter combination in the search for bifurcations, we integrate the climate system from multiple initial conditions and record the temperature extrema after an initial transient interval. Within the parameter range shown, all trajectories remained bounded and the transient evolution decayed sufficiently to reveal the long-term attractor. Each plotted point corresponds to a single post-transient temperature maximum or minimum from an individual simulation. When all initial conditions converge to the same stable equilibrium, the extrema coincide and appear as a single point or narrow curve. In contrast, oscillatory and chaotic solutions produce a broader distribution of extrema, yielding the characteristic vertical spread in the bifurcation diagram.

Thermodynamic phase transitions are often associated with heightened sensitivity and nonlinear behavior, suggesting that complex dynamics might naturally occur near the freezing point of water. However, the present model does not explicitly resolve the microscopic fluctuations associated with a phase transition. Instead, the observed chaos arises from the interaction of climate feedbacks.
Somewhat surprisingly, strong negative feedbacks ($c \lesssim -25$ W m$^{-2}$), in combination with the positive ice-albedo feedback, produce the most complex dynamics in the model. Chaos is concentrated near $T_f \approx 270$--275 K, where the stabilizing fourth feedback overlaps the temperature range over which the ice--albedo feedback is most sensitive. Outside this narrow temperature interval, the climate generally converges to stable states. Importantly, the bifurcation is controlled by the temperature range over which the additional feedback becomes active. Chaos emerges when the climate trajectory repeatedly traverses the temperature interval over which both the stabilizing fourth feedback and the destabilizing ice--albedo feedback are simultaneously active. This condition is most readily satisfied when the fourth feedback transition temperature $T_f$ lies near the ice--albedo transition temperature, causing the two feedbacks to overlap in temperature space. Simply choosing $T_f$ near 273 K is therefore not sufficient to produce chaos; the climate must also evolve through that overlapping region repeatedly so that the competing feedbacks can interact dynamically.

At the largest positive feedback strengths ($c \sim +30$ to $+50$ W m$^{-2}$), the model instead exhibits bistability between warm and snowball climates. In these cases, different initial conditions converge to distinct stable attractors separated by an unstable intermediate state.

\subsection{Volcanism shifts and suppresses chaos}

\begin{figure}[htb!]
    \centering
    \includegraphics[width=0.8\linewidth]{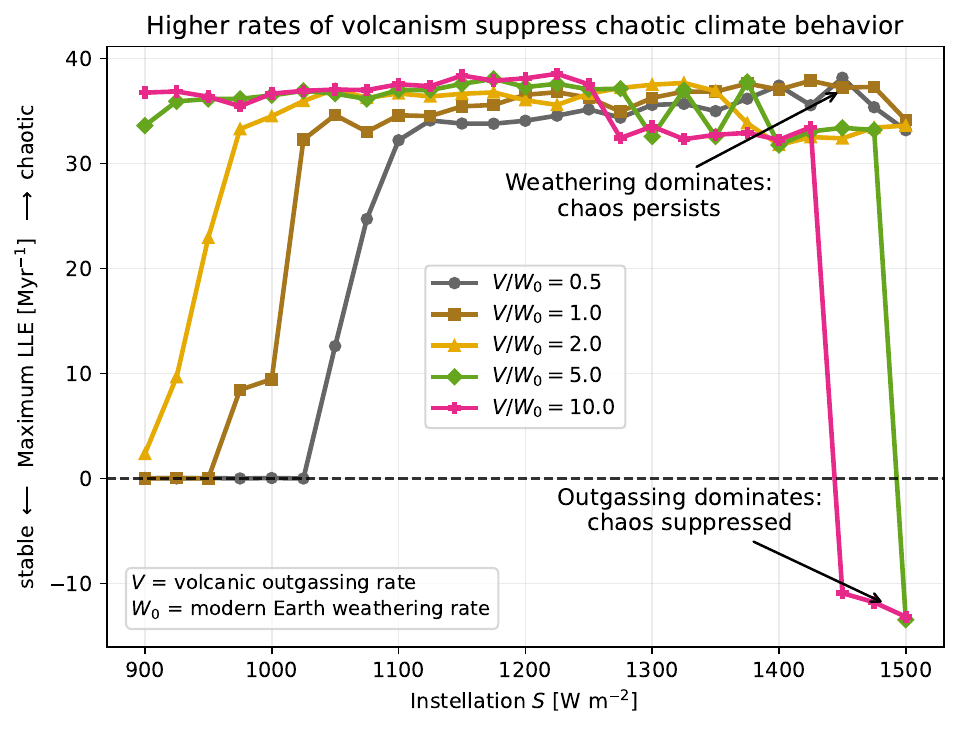}
    \caption{Largest Lyapunov exponent (LLE) as a function of stellar instellation (S) for five values of the normalized volcanic outgassing rate ($V/W_0$). Each curve shows the maximum LLE across the $(c, T_f)$ parameter space at fixed $(S, V/W_0)$, such that positive values indicate the presence of chaotic solutions within that slice. Increasing normalized volcanic outgassing systematically shifts the chaotic regime toward lower stellar flux and suppresses chaos at sufficiently high instellation.}
    \label{fig:volc}
\end{figure}

Figure~\ref{fig:volc} shows how the normalized volcanic outgassing rate modifies the location and intensity of chaotic climate behavior across instellation. For each pair of $(S,V/W_0)$, we plot the maximum LLE obtained across the explored $(c,T_f)$ parameter space, such that positive values indicate the existence of chaotic solutions within that slice. At low to moderate outgassing ($V/W_0 = 0.5$--1$)$, positive LLEs persist across much of the explored instellation range. Increasing $V/W_0$ systematically shifts the chaotic regime toward lower stellar flux and reduces the maximum LLE attained at high instellation. At the largest outgassing rates ($V/W_0 = 10$), the maximum LLE becomes negative for $S \gtrsim 1400~\mathrm{W\,m^{-2}}$, indicating that all sampled configurations converge to stable, non-chaotic attractors. In this high-forcing regime, the climate is driven away from the freezing-adjacent temperature range where the ice--albedo and fourth feedbacks interact most strongly, reducing the opportunity for chaotic dynamics to develop.

\begin{figure*}[htb!]
\centering
\includegraphics[width=0.485\linewidth]{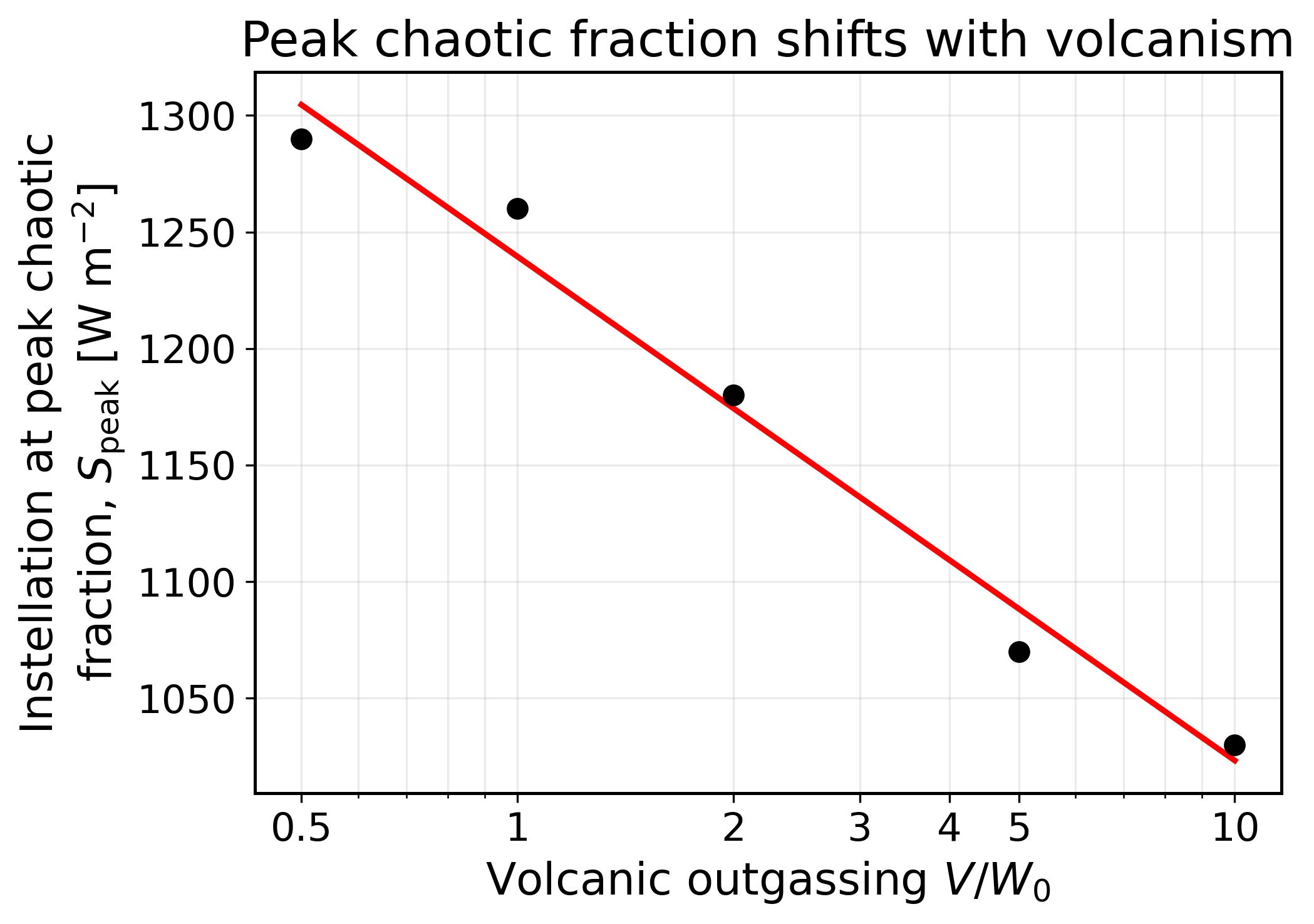}
\hfill
\includegraphics[width=0.507\linewidth]{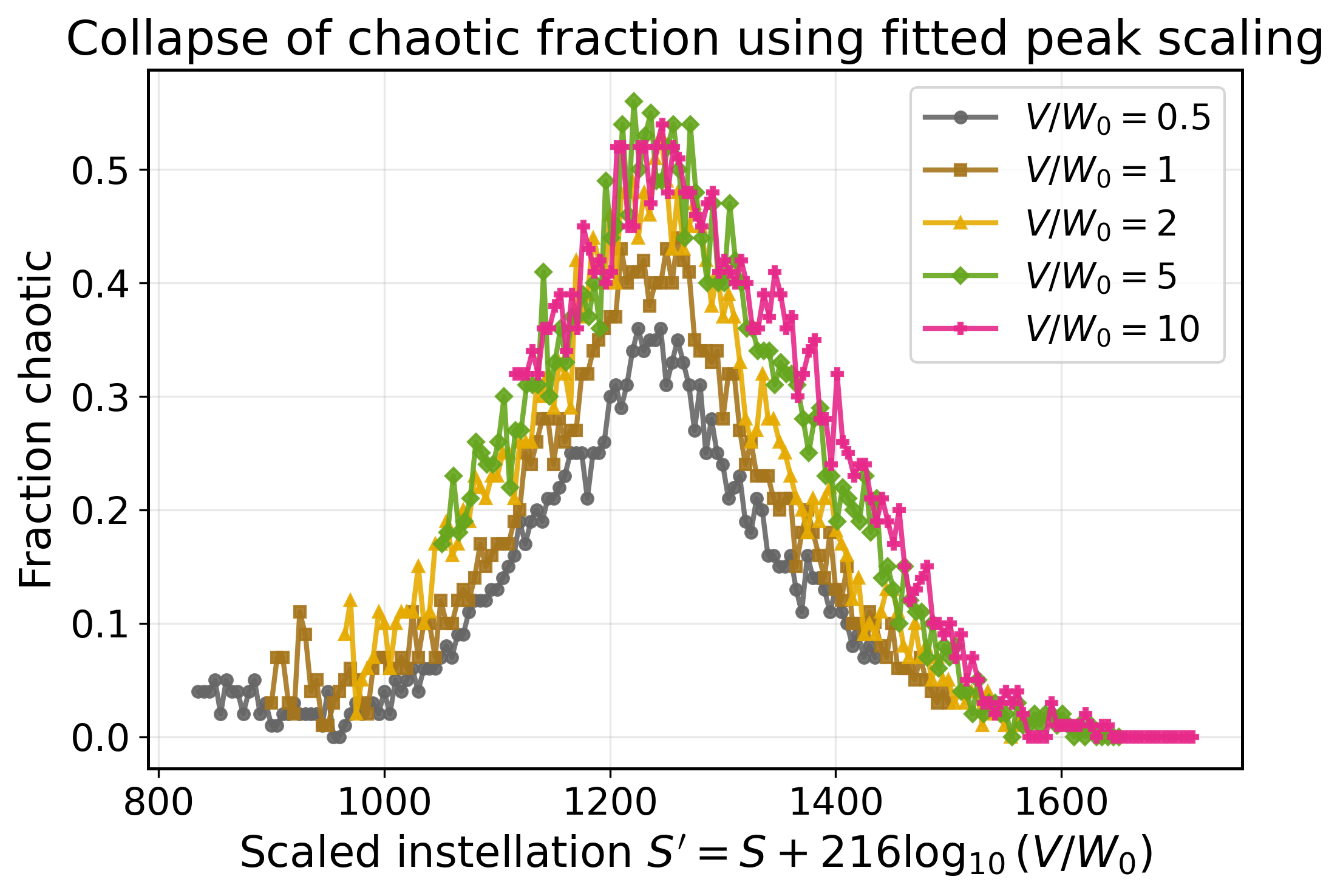}
\caption{
Scaling of the instellation at which the chaotic fraction is maximized for stabilizing fourth feedbacks near the freezing point ($263~{\rm K}<T_f<283~{\rm K}$).
Left: Instellation corresponding to the maximum fraction of chaotic climates, $S_{\rm peak}$, as a function of the normalized volcanic outgassing rate ($V/W_0$), with a logarithmic fit.
Right: Fraction of chaotic climates as a function of shifted instellation, showing that the curves exhibit similar profiles after accounting for the dependence of $S_{\rm peak}$ on $V/W_0$.
Increasing normalized volcanic outgassing shifts the peak of the chaotic regime to lower instellation while preserving the overall shape of the chaotic-fraction curve.
}
\label{fig:volc_scaling}
\end{figure*}

For each normalized outgassing rate, we compute the fraction of parameter combinations that produce chaotic solutions as a function of stellar flux. This chaotic fraction exhibits a bounded, unimodal dependence on instellation (Figure~\ref{fig:volc_scaling}). We define $S_{\rm peak}$ as the stellar flux at which the chaotic fraction reaches its maximum value for a given outgassing rate. Increasing outgassing systematically shifts $S_{\rm peak}$ toward lower instellation, consistent with enhanced greenhouse warming from elevated atmospheric CO$_2$. We find that the location of this maximum is well approximated by 
\begin{equation}
S_{\mathrm{peak}} \approx -216.22 \log_{10}(V/W_0) + 1239.47,
\end{equation}
suggesting that volcanism and instellation combine to define an effective forcing parameter governing proximity to the freezing-adjacent chaotic regime.

When plotted against the shifted coordinate $S + 216\log_{10}(V/W_0)$, the chaotic-fraction curves approximately collapse onto a common profile (Figure~\ref{fig:volc_scaling}). We use the chaotic fraction rather than the maximum LLE for this scaling analysis because it measures the prevalence of chaos across the sampled parameter space, whereas the maximum LLE is determined by a single parameter combination and saturates across much of the chaotic regime; nevertheless, both metrics show that increasing normalized volcanic outgassing ($V/W_0$) shifts chaotic behavior toward lower stellar flux. The collapse is not exact, but the location of the peak chaotic fraction is reproduced with a root mean square error (RMSE) of $14.5~{\rm W\,m^{-2}}$ ($R^2=0.98$), indicating that a large fraction of the dependence on normalized volcanic outgassing can be captured through a simple shift in instellation. The peak chaotic fraction itself varies only modestly across the explored outgassing range, increasing from 0.36 at $V/W_0=0.5$ to 0.56 at $V/W_0=5$ and remaining high (0.54) at $V/W_0=10$. This behavior suggests that volcanism primarily shifts the location of the chaotic regime in parameter space rather than strongly altering its overall prevalence or structure.

\begin{figure}[htb!]
    \centering
    \includegraphics[width=0.8\linewidth]{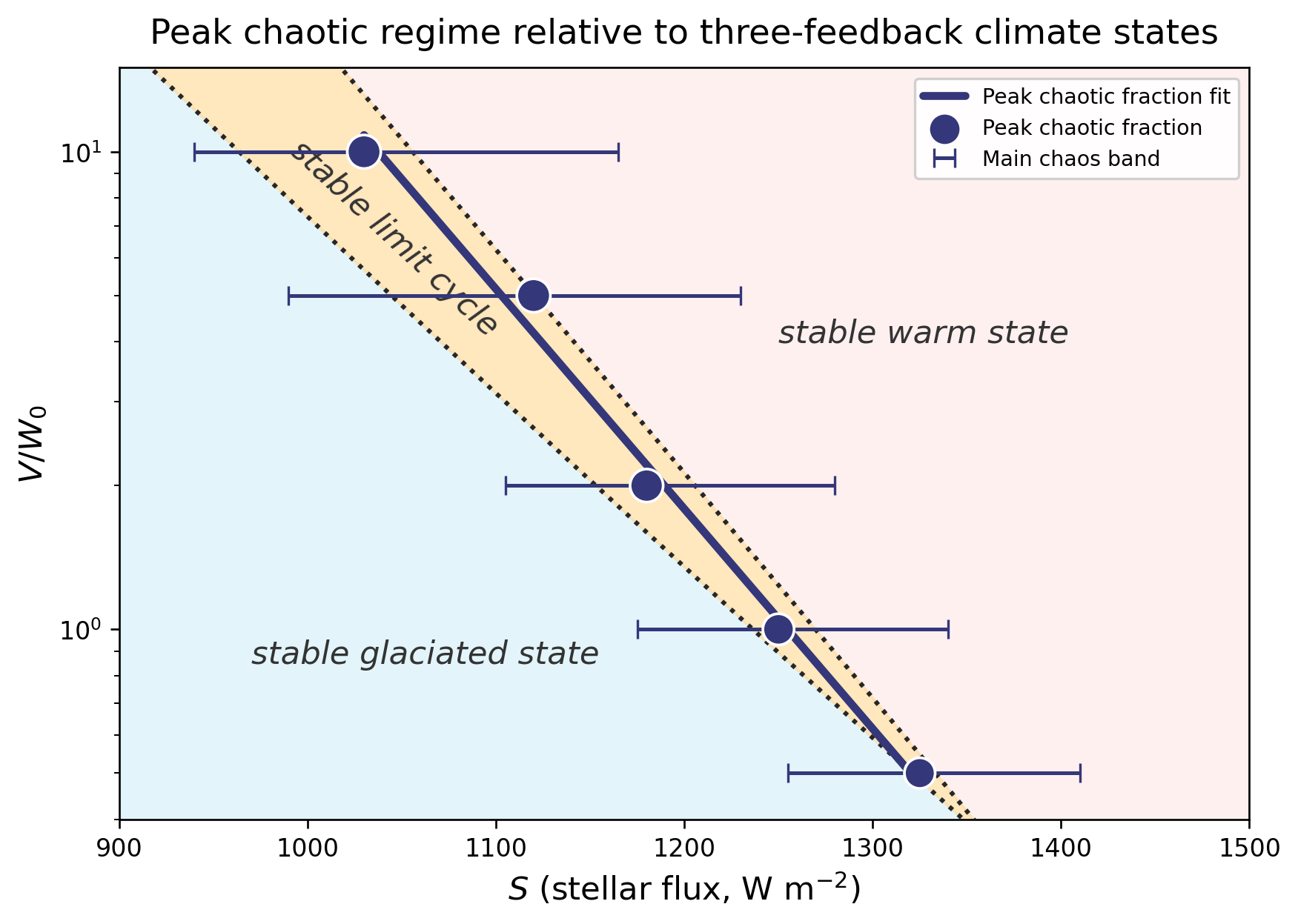}
    \caption{Peak chaotic fraction in $(S, V/W_0)$ space for the four-feedback climate model compared to the behaviors predicted by the three-feedback model. Purple points show the stellar flux at which the chaotic fraction reaches its maximum for each normalized volcanic outgassing rate, while purple horizontal error bars indicate the main chaos band, defined as the range of stellar fluxes where the fraction of chaotic simulations exceeds half of its maximum value for that normalized outgassing rate. The solid line shows the best-fit relation for the peak location. Blue, orange, and red shaded regions show the stable climate regimes predicted by the corresponding three-feedback climate model \citep{Arnscheidt2020} and are included for reference.}
    \label{fig:peakchaos}
\end{figure}

We can compare the distribution of chaotic climates in the four-feedback model to the behaviors predicted in the three-feedback case. For each normalized volcanic outgassing rate, we compute the fraction of parameter combinations that produce chaotic solutions as a function of stellar flux and identify the stellar flux at which this fraction is maximized, $S_{\rm peak}$. Figure~\ref{fig:peakchaos} shows the resulting values of $S_{\rm peak}$ in the ($S,V/W_0$) parameter space. As $V/W_0$ increases, the stellar flux corresponding to the maximum fraction of chaotic climates shifts systematically toward lower values. The relationship is approximately linear in $\log_{10}(V/W_0)$, indicating that stronger volcanic forcing offsets the stellar flux required for climates to remain near the freezing-adjacent transition region where chaotic solutions are most prevalent.

The width of the main chaotic band also varies with volcanic forcing. At low normalized volcanic outgassing, chaotic solutions occur across a broad range of stellar fluxes, whereas at high outgassing rates the chaotic regime narrows and becomes confined to a smaller region of parameter space, behaving similarly to the three-feedback model but with less rapid narrowing. One possible explanation is that the fourth feedback introduces an additional dynamical degree of freedom that allows chaotic attractors to persist over a wider range of forcing than the limit cycles found in the three-feedback model. More generally, adding interacting nonlinear feedbacks may increase the diversity of accessible climate states, although whether this behavior extends to systems with additional feedbacks remains an open question. Outside this band, the system preferentially converges toward either stable glaciated or stable warm climate states. The boundaries between these stable regimes closely track the transition structure predicted by the three-feedback model of \cite{Arnscheidt2020}, with the addition of the fourth feedback producing a finite band of enhanced chaotic behavior centered near the classical habitable transition region.

\section{Discussion}

\subsection{Prevalence and structure of chaotic climate regimes}

Within the finite parameter range and resolution explored here, chaos occupies a minority of the sampled model configurations (10.4\% of simulations). This localization suggests that chaotic behavior is not an automatic consequence of adding a fourth feedback to the climate system. Instead, in our simulations, chaotic behavior is preferentially associated with configurations in which multiple nonlinear feedbacks are simultaneously active over a narrow temperature interval. In particular, the highest occurrence of chaos appears when a stabilizing fourth feedback overlaps the temperature range where the destabilizing ice--albedo feedback is strongest. This result should not be interpreted as a general mathematical condition for chaos in low-dimensional climate systems; rather, it identifies the feedback configurations that favor chaos within the specific four-feedback model and parameter ranges examined here.

The present study holds the ice--albedo transition parameters fixed. Section \ref{Sec:sens_analysis} in the Appendix provides a sensitivity analysis study on the width of the ice--albedo feedback, $T_{\rm scale}$. Future work could investigate how the location and extent of the chaotic regime depend on parameters including albedo contrast, OLR coefficient, and weathering sensitivity, helping to determine whether chaos is fundamentally associated with the freezing point of water itself or, more generally, with the overlap of competing nonlinear feedbacks. A broader sensitivity analysis over these background climate parameters is therefore an important direction for future work.

The concentration of chaos near the freezing point suggests that regions of parameter space where multiple climate feedbacks are simultaneously strong may provide preferred conditions for complex climate dynamics. In the present model, the strongest chaotic behavior occurs when the generalized fourth feedback overlaps the temperature range where the ice--albedo feedback is most sensitive. Near this transition, small temperature perturbations produce large changes in ice cover and planetary albedo, while the fourth feedback acts over a comparable temperature range. The resulting interaction between destabilizing and stabilizing processes operating on similar temperature scales appears to promote chaotic variability. Outside this region, the ice-albedo feedback weakens, the feedbacks become effectively decoupled, and the climate converges to either stable equilibria or regular oscillations.

This concentration has an important implication for planetary climate evolution. If chaotic variability requires the coincidence of multiple threshold-sensitive feedbacks, then intrinsically unpredictable climates may be restricted to specific regions of parameter space rather than being a common feature of terrestrial planets. The prevalence of chaos may therefore depend more strongly on the structure and temperature dependence of planetary feedbacks than on the overall strength of those feedbacks.

\subsection{Feedback interplay as a driver of chaos}

A central result of this work is that chaos emerges when a strong stabilizing feedback activates near the ice--albedo transition ($T_f \approx 270$--282 K). In the model, this behavior arises from the interaction between multiple nonlinear feedbacks operating across a narrow temperature interval where the ice-albedo feedback is strongest. Although the generalized fourth feedback is not tied to a specific physical mechanism, several water-linked processes may plausibly exhibit similar threshold behavior. 

Liquid water is widely regarded as a prerequisite for habitability \citep{Kasting1993} and is expected to drive hydrological cycling on temperate planets \citep{Pierrehumbert2010}. As on Earth, these hydrological cycles are likely coupled to other planetary reservoirs, including the carbon, nitrogen, silicate, and carbonate cycles \citep{Walker1981,Berner1992,Abbot2012}. Because the availability and transport of liquid water change rapidly across the freezing transition, feedbacks associated with these cycles may naturally attain their greatest sensitivity near the ice--liquid boundary \citep{Pierrehumbert2005}. Consequently, strong positive or negative climate feedbacks that overlap the ice-albedo transition may be common on habitable worlds, making freezing-adjacent chaotic climate dynamics a potentially generic outcome rather than a peculiarity of the specific feedback considered here. 

We emphasize that the chaotic regime identified here is associated with the freezing transition itself rather than the outer edge of the habitable zone. Depending on volcanic outgassing and atmospheric CO$_2$, planets spanning a range of stellar fluxes may repeatedly traverse this temperature interval, whereas climates near the classical inner habitable-zone boundary generally remain well above the regime where the ice–albedo feedback is active.

This result highlights the importance of where in temperature space a feedback operates, not just its magnitude or sign. Feedbacks that are stabilizing in isolation can induce complex, even chaotic, dynamics when they act in the vicinity of sharp nonlinearities such as phase transitions. 

\subsection{Volcanic outgassing shifts and partially suppresses chaos}

Volcanic outgassing primarily shifts the location of chaotic regimes in parameter space while partially suppressing their strength at high $V/W_0$. Increasing $V/W_0$ systematically moves the peak chaotic fraction toward lower stellar flux, consistent with stronger greenhouse warming from elevated atmospheric CO$_2$. When plotted against the shifted coordinate $S+\alpha\log V$, the chaotic-fraction curves approximately collapse onto a common profile, indicating that volcanism acts largely as an effective forcing offset rather than fundamentally altering the underlying bifurcation structure.

At the highest normalized outgassing rates, however, the maximum LLE and peak chaotic fraction both decrease, suggesting partial stabilization of the climate system. Physically, stronger outgassing shifts the baseline climate away from repeated crossings of the ice--albedo transition that are associated with chaotic variability in the model. Together, these results suggest that volcanism controls both the location and accessibility of freezing-adjacent chaotic regimes.

\subsection{Potential Physical Origins of a Fourth Feedback}

The fourth feedback modeled in this study is intended to serve as a proxy for climate mechanisms that are not included in the three-feedback models of terrestrial climate used in previous studies. The class of fourth feedbacks that produce the chaotic behavior identified in this study share three properties: (1) a stabilizing influence on climate, (2) strongly nonlinear behavior, and (3) activation near the freezing point of water. While the ice--albedo feedback also satisfies some of these properties, this fourth feedback differs in that it acts to restore the climate toward an equilibrium state rather than amplify temperature perturbations. Thus, when the two feedbacks operate over similar temperature ranges, they compete: the ice--albedo feedback destabilizes the climate by reinforcing departures from equilibrium, whereas the fourth feedback counteracts those departures. The chaotic behavior identified in this study appears to arise from the interaction of these competing nonlinear feedbacks near the freezing transition.

\begin{table*}[htb!]
\centering
\small
\begin{tabular}{l c c c}
\hline
\textbf{Candidate fourth feedback} &
\textbf{Stabilizing tendency} &
\textbf{Potential forcing influence} &
\textbf{Near Freezing} \\
\hline

\multicolumn{4}{c}{\textbf{Water-linked candidates}} \\
\hline

Seafloor weathering
& Moderate--High
& Moderate
& Moderate \\

Ocean alkalinity--CO$_2$ partitioning
& High
& Moderate--High
& Moderate--High \\

Ocean circulation / stratification
& Moderate
& Moderate
& High \\

Pore-space carbonate precipitation
& Moderate
& Moderate
& Moderate \\

CH$_4$ clathrate stability
& Destabilizing
& Moderate--High
& Moderate--High \\

CO$_2$ ocean solubility
& Mixed
& Moderate
& Moderate \\

Hydrothermal CO$_2$ exchange
& Unclear
& Low--Moderate
& Low--Moderate \\

\hline
\multicolumn{4}{c}{\textbf{Atmospheric and surface feedbacks}} \\
\hline

Cloud radiative feedback
& Mixed
& High
& Low \\

Water vapor greenhouse feedback
& Destabilizing
& High
& Low \\

CH$_4$--SO$_2$ photochemical haze
& Moderate
& Moderate
& Low \\

Dust / aerosol radiative feedback
& Moderate
& Moderate
& Low \\

Sulfur surface deposition / albedo
& Moderate
& Low
& Low--Moderate \\

Temperature-dependent serpentinization
& Destabilizing
& Low
& Low \\

\hline
\end{tabular}

\caption{
Candidate climate feedbacks that may satisfy some of the dynamical conditions associated with chaotic behavior in our model, including nonlinear temperature dependence, partial stabilization, and enhanced sensitivity near the freezing point of water. The qualitative classifications (``high,'' ``moderate,'' and ``low'') reflect our heuristic assessment of the relative stabilizing influence, potential climate forcing, and expected sensitivity near the freezing point based on published physical understanding, rather than quantitative model predictions or definitive categorizations. Feedbacks involving ocean circulation, sea ice, and liquid-water availability are among the most plausible freezing-adjacent candidates.
}
\label{tab:fourth_feedback_candidates}
\end{table*}

Among the candidate mechanisms summarized in Table~\ref{tab:fourth_feedback_candidates}, ocean-mediated feedbacks appear most consistent with these requirements. Ocean alkalinity--CO$_2$ partitioning regulates the distribution of carbon between the ocean and atmosphere, while ocean circulation controls the storage and ventilation of dissolved carbon reservoirs \citep{Walker1981,Berner2004,Toggweiler1999,Sigman2010}. Both processes can exert substantial influence on atmospheric CO$_2$ and may become strongly nonlinear near the ice--liquid transition as sea-ice coverage, stratification, and overturning circulation reorganize.

Ocean circulation feedbacks may be particularly attractive because they provide a natural explanation for why chaotic behavior is concentrated near the freezing point. Sea-ice formation and melting alter salinity through brine rejection, modifying deep-ocean ventilation and carbon storage. Because sea-ice fraction changes rapidly near 273 K, circulation-driven carbon cycle feedbacks naturally become strongest in the same temperature range where chaos emerges in the model \citep{Toggweiler1999,Sigman2010,Rahmstorf2002}.

Planetary obliquity may likewise alter the effective strength and activation temperature of the generalized fourth feedback. By changing the seasonal distribution of insolation, obliquity can modify sea-ice formation, ocean circulation, and climate hysteresis \citep{Olson2020}. This suggests that the location and extent of chaotic climate regimes may depend not only on atmospheric composition and stellar flux, but also on a planet's rotational state. Incorporating obliquity into reduced-order climate models therefore represents a promising direction for future work.

Hydrologically mediated geochemical feedbacks, including seafloor weathering and ocean alkalinity evolution, provide plausible alternatives. These mechanisms can stabilize climate through long-term carbon sequestration and may be particularly important on ocean worlds lacking extensive continental weathering \citep{Abbot2012,KrissansenTotton2018,Pierrehumbert2010,Graham2020}. Because seafloor weathering depends on ocean circulation to transport dissolved CO$_2$ to reactive basaltic crust, changes in ocean circulation near the freezing transition may also enhance its temperature sensitivity, making it another plausible candidate for the generalized fourth feedback.

Many classical atmospheric feedbacks appear less consistent with the behavior identified in our simulations. Water-vapor, cloud, aerosol, and photochemical feedbacks can strongly affect climate, but generally operate across broad temperature ranges rather than activating sharply near the liquid--ice transition \citep{Soden2006,Sherwood2020}.

\subsection{Predictability limits and implications for orbital forcing}

The maximum LLE values reach $\sim$30--40 Myr$^{-1}$, corresponding to Lyapunov timescales of $\sim$25--33 kyr. The Lyapunov timescale represents the characteristic time over which initially nearby climate trajectories diverge by a factor of $e$, providing a measure of the intrinsic predictability horizon of the system. This implies that uncertainties in initial conditions are largely amplified and the detailed evolution of the climate becomes increasingly difficult to predict beyond several tens of thousands of years, even though the climate remains confined to a bounded region of phase space.

This overlap has important implications. When internal variability and external forcing occur on comparable timescales, their interaction may be nontrivial, potentially leading to phase locking, modulation, or amplification of climate variability. As a result, even well-characterized orbital parameters may not yield straightforward predictions of long-term climate evolution for planets in this regime.

More broadly, these findings suggest that the concept of a single, stable habitable climate state may be insufficient for some exoplanets. Instead, planets near critical feedback thresholds may experience inherently unpredictable climate trajectories, with implications for both long-term habitability and the interpretation of observational data.

\section{Conclusions} \label{sec:conclusions}

We systematically explored the dynamical behavior of a generalized four-feedback climate model across 35,864 parameter combinations spanning stellar flux, volcanic outgassing, feedback strength, and feedback activation temperature. Using the largest Lyapunov exponent as a diagnostic, we identified regions of parameter space that produce chaotic climate evolution and characterized how these regions depend on planetary forcing and feedback properties. 

We find that chaotic dynamics emerge preferentially when a strong stabilizing feedback activates over the same temperature range in which the ice--albedo feedback is most sensitive. In the present model, this overlap occurs near the freezing transition of water ($T_f \approx 270$--282 K), where small temperature perturbations produce large changes in planetary ice coverage and albedo. Within this region, chaotic solutions are enriched by a factor of $\sim$2.5 relative to the full parameter space. These results suggest that chaos is not associated with the freezing point itself, but rather with the interaction of multiple nonlinear feedbacks operating simultaneously over a narrow temperature interval.

We also find that increasing the normalized volcanic outgassing rate $V/W_0$ systematically shifts the region of maximum chaotic occurrence toward lower stellar fluxes. The stellar flux corresponding to the maximum fraction of chaotic climates follows an approximately logarithmic dependence on outgassing rate, indicating that volcanic forcing and stellar flux combine to define an effective control parameter governing proximity to the chaotic regime. Despite this shift, the overall structure of the chaotic region remains largely unchanged.

The strongest chaotic solutions exhibit largest Lyapunov exponents of $\sim$30$-$40 Myr$^{-1}$, corresponding to Lyapunov timescales of roughly 25--33 kyr. These timescales represent the characteristic horizon over which nearby climate trajectories diverge, implying that the detailed evolution of such climates may become intrinsically unpredictable on timescales of tens of thousands of years.

Thus, if similar (fourth) feedbacks operate on terrestrial exoplanets, climates near phase transitions may be intrinsically unpredictable rather than simply bistable. This raises the possibility that some planets within the nominal habitable zone experience persistent, chaotic climate variability, complicating both habitability assessments and observational interpretation. Future work incorporating potential fourth feedbacks, especially if strongly stabilizing near the freezing point of water, will be critical for determining how widespread such regimes may be. 

\begin{acknowledgments}
    We thank R. Pierrehumbert, R. Wordsworth, T. Robinson, M. Marley, S. Ranjan, T. Barman, and A. Salazar for helpful and constructive comments.
    This material is based upon work supported by the National Aeronautics and Space Administration under Agreement No. 80NSSC21K0593 for the program ``Alien Earths". This publication was partly funded by the Heising-Simons Foundation through grant \#2024-5688.

    This material is based upon High Performance Computing (HPC) resources supported by the University of Arizona TRIF, UITS, and Research, Innovation, and Impact (RII) and maintained by the UArizona Research Technologies department. This research has made use of NASA’s Astrophysics Data System.
\end{acknowledgments}

\begin{contribution}
C.L. and D.A. designed the research project, and planned the structure of the paper and figures. D.A. secured funding for the project. C.L. developed model, carried out analysis, made the figures, and drafted the manuscript. C.L. and D.A. worked jointly on the interpretation.  D.A. and R.M. provided comments on the manuscript.

\end{contribution}


\appendix

\section{Largest Lyapunov Exponent Grid}

\begin{figure}[htb!]
\centering
\includegraphics[width=\linewidth]{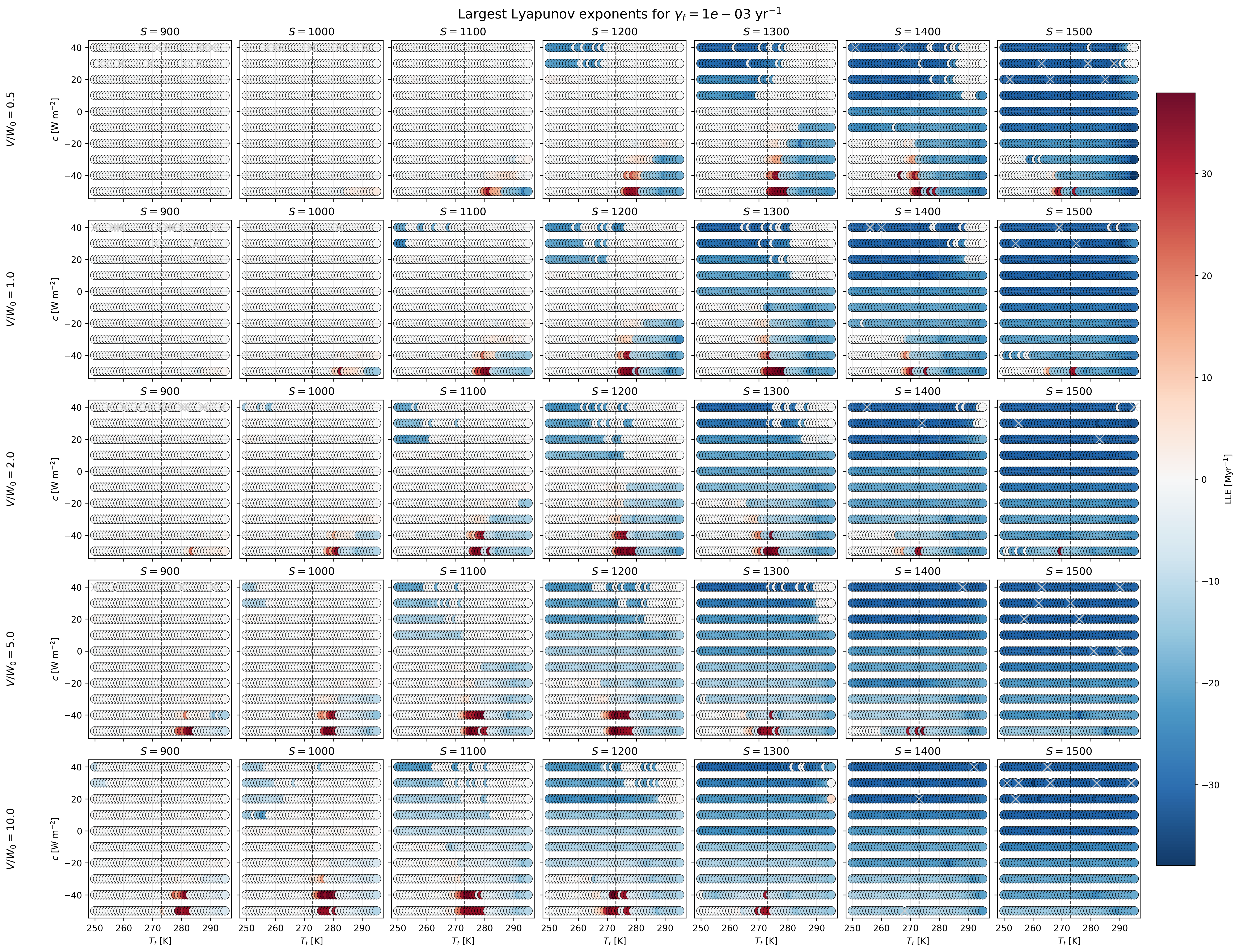}
\caption{Largest Lyapunov exponent (LLE) across the four-feedback parameter space as a function of feedback strength $c$ and activation temperature $T_f$. Columns show increasing stellar flux and rows show increasing normalized volcanic outgassing rate $(V/W_0)$. Warm colors indicate chaotic dynamics (positive LLE), while cool colors indicate regular attractors (negative LLE). The vertical dashed line marks the freezing point of water (273 K). Chaos is concentrated where a strong stabilizing feedback ($c \lesssim -30$ W m$^{-2}$) activates near the ice--albedo transition, with the region of maximum chaos shifting toward lower stellar flux as volcanic outgassing increases.}
\label{fig:grid search}
\end{figure}

Figure~\ref{fig:grid search} presents the full parameter survey used in this study. Each panel shows the largest Lyapunov exponent (LLE) as a function of feedback strength $c$ and transition temperature $T_f$ for a particular combination of stellar flux $S$ and volcanic outgassing rate $V/W_0$. Positive LLE values indicate chaotic dynamics, while negative values correspond to regular attractors (stable equilibria or limit cycles).

The full survey reveals that chaotic solutions occupy a localized but coherent region of parameter space. Chaos is strongly concentrated near the freezing point of water and becomes most prevalent when the fourth feedback is strongly stabilizing. As volcanic outgassing increases, the stellar flux corresponding to the highest fraction of chaotic solutions systematically decreases, producing a diagonal ridge of chaos across the $(S,V)$ parameter space.

\section{Sensitivity of Chaos to the Width of the Ice--Albedo Transition}
\label{Sec:sens_analysis}

To assess the sensitivity of our results to the width of the ice--albedo transition, we varied $T_{\rm scale}$ while holding all other background climate parameters fixed. This limited sensitivity test is intended to evaluate whether the localization of chaos near the freezing transition is robust to modest changes in the ice--albedo transition width, rather than to provide a comprehensive exploration of the full parameter space.

\begin{figure}[htb!]
    \centering
    \includegraphics[width=\linewidth]{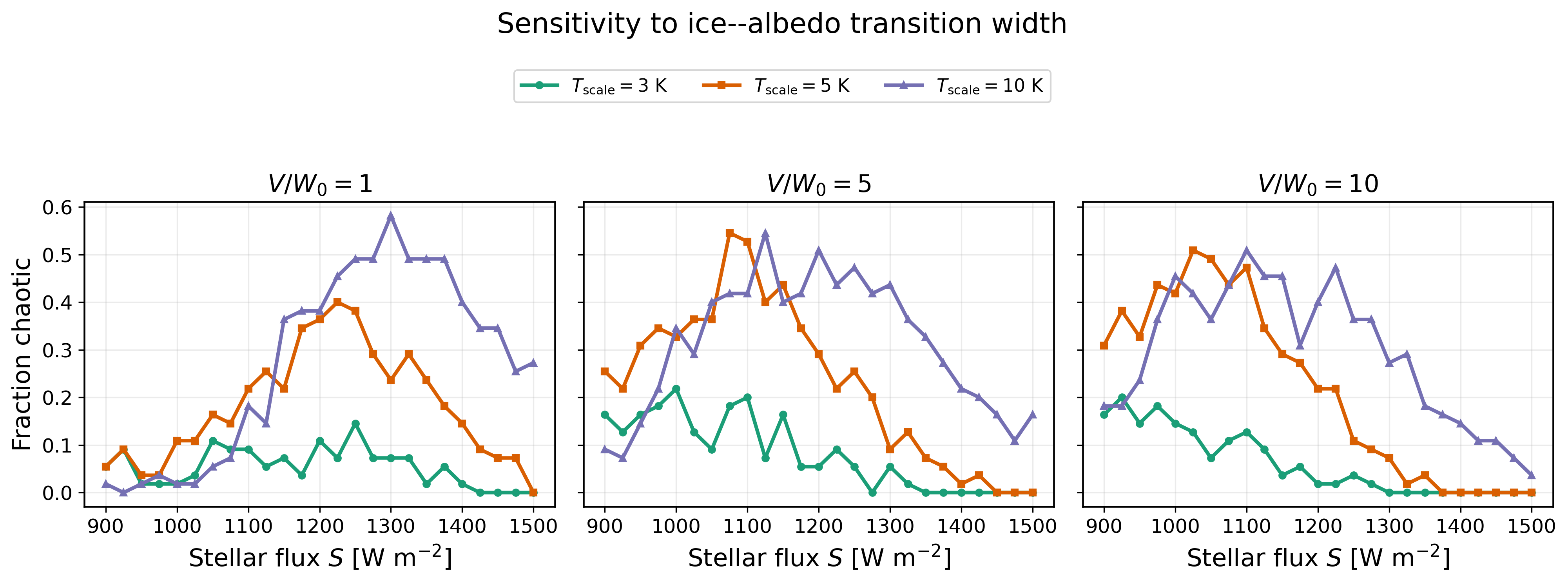}
    \caption{Sensitivity of the chaotic fraction to the ice--albedo transition width $T_{\rm scale}$. Results are shown for three representative outgassing rates. Broader ice--albedo transition widths increase both the prevalence and parameter-space extent of chaos while preserving the same overall dependence on  stellar flux.}
    \label{fig:tscale_sensitivity}
\end{figure}

Figure~\ref{fig:tscale_sensitivity} shows the fraction of chaotic climates as a function of stellar flux for $T_{\rm scale}=3$, 5, and 10~K and three representative volcanic outgassing rates. Increasing $T_{\rm scale}$ systematically increases both the peak chaotic fraction and the range of stellar flux over which chaotic climates occur. In contrast, the location of the peak chaotic fraction depends much more strongly on volcanic outgassing than on $T_{\rm scale}$, continuing to shift toward lower stellar flux as volcanic outgassing increases. Thus, while the prevalence of chaos depends on the sharpness of the ice--albedo transition, the principal conclusions of this work remain unchanged: chaotic climates remain concentrated near the freezing transition, and their location in parameter space is primarily controlled by volcanic outgassing.

\bibliography{sample701}{}
\bibliographystyle{aasjournalv7}

\end{document}